\documentclass[aps,superscriptaddress,twocolumn,fleqn,showpacs]{revtex4-1}
\usepackage{here}
\usepackage{amsmath,amssymb}
\usepackage{mathrsfs}
\usepackage{bm}
\usepackage[dvipdfmx]{graphicx}
\usepackage[dvipdfmx]{color}

\usepackage[normalem]{ulem}

\newcommand{\FIG}[1]{{Fig.~#1}}
\newcommand{\CITE}[1]{{\cite{#1}}}

\def\eps{\varepsilon}

\begin{document}

\title{A Simple Generic Model of Cellular Polarity Alignment:\\
Derivation and Analysis
}
\author{Kaori Sugimura}
\affiliation{Department of Information Sciences, Ochanomizu University, Tokyo 112-8610, Japan}
\author{Hiroshi Kori}
\email[corresponding author: ]{kori.hiroshi@is.ocha.ac.jp}
\affiliation{Department of Information Sciences, Ochanomizu University, Tokyo 112-8610, Japan}

\date{\today}

\begin{abstract}
 Ordered polarity alignment of a cell population plays a vital role in
 biology, such as in hair follicle alignment and asymmetric cell division.
 Here, we propose a theoretical framework for the understanding of
 generic dynamical
 properties of polarity alignment in
 interacting cellular units, where each cell
 is described by a reaction-diffusion system and 
 the cells further interact with one another through their proximal surfaces.
 The system behavior is shown to be strongly dependent on geometric
 properties such as cell alignment and cell shape. 
 Using a perturbative method under the assumption of weak coupling between cells,
 we derive a reduced model in which each cell is described by just one
 variable, the phase. 
 The reduced model resembles an XY model but contains novel terms
 that possesses geometric information, which
 enables the understanding of the geometric dependencies as well
 as the effects of external signal and noise.
 The model is simple, generic, and analytically and numerically
 tractable, and is therefore expected to facilitates
 studies on cellular polarity alignment in various nonequilibrium systems.
\end{abstract}

\maketitle
{\em Introduction--.}
Spatially ordered patterns are ubiquitous in nature and
have been of central importance in various disciplines
\CITE{cross93, meinhardt2000pattern, nakao2010turing}. 
This work is concerned with
dynamical alignment of polarity in interacting cellular units.
Spin is a prototypical example of a polar unit. Spins are
spatially aligned to magnetize
through spin-spin interaction and their response to an external field,
as described by, e.g., Ising and XY models \cite{kosterlitz1974critical}.
The present work focuses on nonequilibrium systems, including chemical
and biological systems. Polarity can be regarded as
an asymmetric distributions in chemical species within a cellular
unit. 
Polarity is of great importance in biology because
it is essential for, e.g., cell movement and oriented cell division \cite{devenport2014cell}.
Moreover, cell polarity is often spatially coordinated across a cell
population for functional reasons.
A well-known example in biology is planar cell polarity (PCP),
which refers to the coordinated alignment of cell polarity across planar tissue.
This underlies the alignment of, e.g., hair follicles and cilia positioning
\cite{devenport2014cell}.
So far, several mathematical models have been proposed to address
the effects of various factors on polarity alignment including cell
shape, external signal, and noise. 
Some studies employ detailed models, where each cell is
described by a reaction-diffusion system and these cells are further coupled
through proximal membranes \CITE{amonlirdviman05, burak09}.
Some studies employ simple phenomenological models
similar to models for magnetization\CITE{aigouy10, ayukawa2014dachsous},
which is a reasonable approach because the cell
alignment process phenomenologically resembles magnetization.
The former contains many free parameters and
is too complicated to provide general understandings. 
On the other hand, in the latter, models are rather arbitrary
and may lack essential dynamical features.

In the present Letter,
the generic dynamical properties of cell polarity alignment are
examined, through deriving a
reduced model for coupled reaction-diffusion systems
using a perturbative method.
In our reaction-diffusion model, each cell
is described by a reaction-diffusion system and 
the cells mutually inhibit one another through their proximal surfaces.
Its reduced model, referred to as a phase model,
is drastically simple yet reasonably approximates the original
reaction-diffusion model when cells are weakly coupled.
Our phase model resembles the XY
model but includes novel terms representing geometric information
such as cell shape and the relative position between neighboring cells.
By taking advantage of its tractability,
essential dynamical properties including the effects
of cell shape, external signal, and noise are analytically clarified,
which has only been studied
numerically in previous studies using detailed models
\cite{amonlirdviman05, burak09, aigouy10}.
Our study bridges the gap between detailed and phenomenological models,
and is expected to facilitate the study of polarity dynamics in various
nonequilibrium systems.

\begin{figure}[t]
  \includegraphics[scale=1.0]{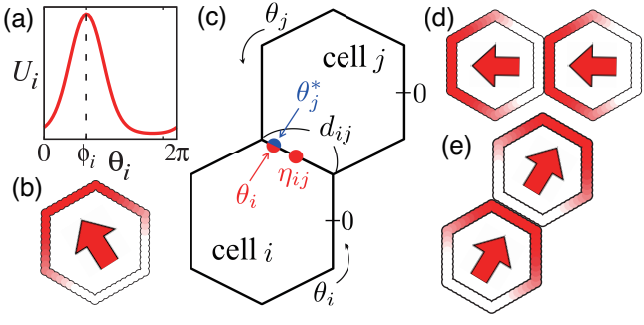}
  \caption{(Color online).
  (a) Unimodal distribution on the surface of a cell and polarity
  orientation. (b) Color scale representation and polarity
  orientation corresponding to (a). (c) Model description. (d,e)
  Examples of polarity patterns of two coupled cells with
 different cell alignments.}
 \label{fig:illust}
\end{figure}

{\em Model--.}
Our entire system is composed of a population of cells 
aligned in two-dimensional space.
Reaction-diffusion dynamics of each cell
take place on the one-dimensional surface of their perimeter $2\pi$, and the 
cells further interact with one another 
through the proximal surfaces between them.
Each cell obeys
\begin{align}
 \frac{\partial}{\partial t} \bm{X}_{i}
  &= \bm{F}(\bm{X}_{i})
  + \hat D \frac{\partial^2 \bm{X}_{i}}{\partial \theta_i^2}
 + \epsilon \sum_{j\in A(i)} \bm H_{ij},
\label{noudo}
\end{align}
where $\bm{X}_i = \bm{X}_i(\theta_i, t)$ ($i=1,\ldots,N$) denotes
the concentration of
chemical species at 
time $t$ and the position $\theta_i$ ($0\le \theta_i < 2\pi$)
on the surface of each cell,
$\bm{F}$ describes the local reaction dynamics, 
$\hat D$ is the diagonal matrix consisting of diffusion coefficients,
$A(i)$ is the set of cells adjacent to cell $i$, 
$\bm{H}_{ij}$ describes intercellular interaction,
and $\eps$ is the coupling strength.
As will be described later, external signal and noise may also be considered.
Note that $\bm{H}_{ij}$ is generally a functional of $\bm
X_i(\theta_i, \cdot)$ and $\bm X_j(\theta_j, \cdot)$.
Each cell is assumed to exhibit a unimodal distribution for $\eps=0$; i.e.,
polarity is spontaneously formed.
The polarity orientation of cell $i$ is defined
by the $\theta_i$ value at which
the first component of $\bm X_i(\theta_i, t)$, denoted by $U_i(\theta_i,
t)$,
takes its maximum [see Figs.~\ref{fig:illust}(a,b)].

 As examples, we consider
two models: (a) the real Ginzburg-Landau equation
(GLE) and (b) the activator-inhibitor model.
Both of these models have two variables, denoted by $\bm X_i = (U_i, V_i)$
(see Supplemental Material A for details).
The former is a long-wave amplitude equation, which is widely used to describe
various systems near the onset of instability.
The latter is a
reaction-diffusion model, describing biological pattern formation \CITE{koch94}.
In these models, given appropriate initial conditions,
$\bm X_i$ exhibits a stationary unimodal distribution
within individual cells for $\eps=0$, 
thus they are suitable as dynamical models describing cell polarity.

Intercellular interaction is assumed to occur
at every contact point of the neighboring cells.
Geometric parameters are defined as shown in
Fig.~\ref{fig:illust}(c), where $\eta_{ij}$ and $d_{ij}$ 
are the midpoint and the length of the proximal surface between cell $i$ and
$j$, respectively, and $\theta_j^*$ is the position in cell $j$ facing
$\theta_i$ in cell $i$. 
The interaction function is given as
\begin{align}
 \bm{H}_{ij} = (S_{ij}(\theta_i-\eta_{ij}) \left\{ U_i(\theta_i) - U_j (\theta_j^*)
 \right\}, 0).
 \label{Hij}
\end{align}
where
$S_{ij}(\theta_i-\eta_{ij})$ desribes the position of contact, given as
$S_{ij}(\theta)=1$ for $|\theta| \leq \frac{d_{ij}}{2}$ and
$S_{ij}(\theta)=0$ otherwise.
When the cell has a regular
hexagonal shape, which is assumed henceforth unless otherwise noted,
we have $d_{ij}=\frac{\pi}{3}$ and $\theta_j^* = \pi + 2 \eta_{ij} - \theta_j$.
The latter relationship holds true also for elongated hexagonal shapes
introduced later.
For $\epsilon>0$, Eq.~\eqref{Hij} describes mutual inhibition of $U$
component between proximal cells. 
With this interaction, the polarities of two neighboring cells are
expected to align along their relative position of the cells, i.e.,
$\eta_{ij}$ or $\pi+\eta_{ij}$,
because the surface of cell $i$ with high $U$ tends to face
that of cell $j$ with low $U$, 
as is illustrated in Figs.~\ref{fig:illust}(d) and (e).

\begin{figure}[t]
 \begin{center}
  \includegraphics[scale=1.0]{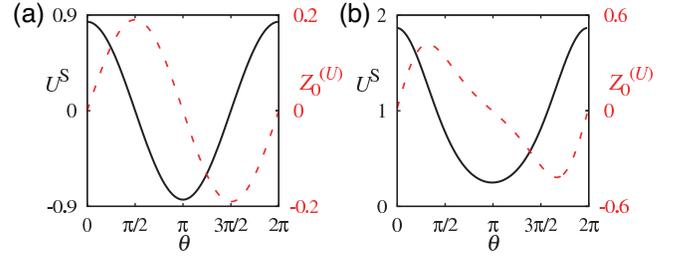}
 \end{center}
 \caption{(Color online). The profile of the steady state $U^{\rm
 S}(\theta)$ (solid lines)
  and the phase sensitivity function $Z_0^{(U)}(\theta)$ (dashed lines)
 for (a) the GLE
 and (b) the activator-inhibitor model. }
   \label{fig:X_Z}
\end{figure}

{\em Derivation of the phase model--.} 
We derive a reduced model for Eq.~\eqref{noudo}
using a perturbative method.
Our method is based on well-known phase reduction theory
\cite{kuramoto84}
and is an application of the recently developed method for oscillatory
patterns reported in Refs. \cite{nakao14,kawamura15}.

Let $\bm X^{\rm S}(\theta)$ be the stationary distribution of a cell
in the unperturbed system ($\eps=0$).
Because of the translational symmetry,
$\bm X^{\rm S}(\theta-\theta_0)$ with any constant $\theta_0$ is also a steady solution.
The phase $\phi_i(t)$ of the $\bm{X}_i(\theta_i,t)$ is defined such that
$\bm{X}_i(\theta_i,t)$ converges to $\bm{X}^{\rm S}(\theta_i-\phi)$ as
$t\to \infty$ in the unperturbed system. In other words,
$\bm y_i(\theta_i,t) \to 0$ as $t \to \infty$ for $\epsilon=0$, 
where the deviation $\bm y_i(\theta_i,t)$ is defined by
\begin{equation}
 \bm X_i(\theta_i,t) =  \bm X^{\rm S}(\theta_i-\phi_i) + \bm y_i(\theta_i,t),
  \label{deviation}
\end{equation}
with $\phi_i$ being the phase of the state $\bm X_i(\theta_i,t)$.
Without loss of generality, we assume that $U^{\rm S}(\theta)$, which is
the $U$ component of $\bm X^{\rm S}(\theta)$, takes
its maximum at $\theta=0$. Then, for sufficiently small $\bm y_i(\theta_i,t)$,
$\phi_i(t)$ of $\bm X_i(\theta_i,t)$ is well approximated by the maximum of
$U_i(\theta_i,t)$, i.e.,
\begin{equation}
 \phi_i(t) \approx {\rm argmax}_{\theta_i} U_i(\theta_i, t).
  \label{max}
\end{equation}
Thus, $\phi_i$ may be regarded as the polarity orientation of cell $i$.

The linear operator $\mathcal L$ is defined by
$\mathcal L = J + \hat D \frac{\partial^2}{\partial \theta^2}$
with Jacobian $J=\partial \bm F(\bm X)/\partial
\bm X$ estimated at $\bm X = \bm X^{\rm S}(\theta)$.
The adjoint operator $\mathcal L^\dagger$ is defined such that it
satisfies $\langle \bm A, \mathcal L \bm B \rangle = \langle \mathcal L^\dagger
\bm A, \bm B \rangle$, where 
the inner product of the $2\pi$-periodic functions, $\bm A(\theta)$
and $\bm B(\theta)$, is defined by
$\langle \bm A, \bm B \rangle = \int_0^{2\pi} \bm A \cdot \bm B d \theta$.
For our model \eqref{noudo}, we can show that
$\mathcal L^\dagger = J^{\rm T} + \hat D\frac{\partial^2}{\partial
\theta^2}$, where $J^{\rm T}$ is the transpose of $J$.
The eigenfuncitions of $\mathcal L$ and $\mathcal
L^\dagger$ are denoted by $\bm Y_\ell(\theta)$ and $\bm Z_\ell(\theta)$
($\ell=0,1,\ldots$), respectively. In particular, the zero-eigenfunctions are
denoted by $\bm Y_0$ and $\bm Z_0$, i.e., $\mathcal L \bm Y_0 = \mathcal
L^\dagger \bm Z_0 = 0$.
Here, we choose $\bm Y_0 = -\frac{\partial \bm X^{\rm S}}{\partial
\theta}$. These eingenfunctions are assumed to form a complete
orthonomal system
and are normalized as $\langle \bm Z_\ell, \bm Y_m \rangle =
\delta_{\ell m}$. The deviation $\bm y_i$ can be expanded as
\begin{equation}
 \bm y_i(\theta_i,t) = \sum_{\ell=1}^\infty \bm C_\ell(t)
  \bm Y_\ell(\theta_i - \phi_i),
  \label{expansion}
\end{equation}
where $\phi_i$ is the phase of the state $\bm X_i(\theta_i,t)$.
Note that $\bm Y_0(\theta_i-\phi_i)$ is absent in this expansion because $\bm y_i(\theta_i,t) \to
0$ as $t\to \infty$ for $\eps=0$.

Substituting Eq.~\eqref{deviation} into Eq.~\eqref{noudo}, we obtain
\begin{equation}
 \bm Y_0(\theta_i-\phi_i) \dot \phi_i + \dot {\bm y}_i = \mathcal L \bm y_i
  + \eps \sum_{j\in A(i)} \bm H_{ij} + O(\eps^2).
\end{equation}
Taking the inner product with $\bm Z_0 (\theta_i-\phi_i)$ and dropping $O(\eps^2)$,
we finally obtain the phase model given as
\begin{align}
 \dot \phi_i = \eps \sum_{j\in A(i)} \Gamma_{ij} (\phi_i,\phi_j)
 \label{phase_model},\\
 \Gamma_{ij} = \langle \bm Z_0 (\theta_i-\phi_i),
 \bm H_{ij}^{\rm S} \rangle,
 \label{gamma}
\end{align}
where $\bm H_{ij}^{\rm S} = \bm H_{ij}\left\{\bm X^{\rm S}(\theta_i -
\phi_{i}), \bm X^{\rm S}(\theta_j - \phi_{j})\right\}$.
Given the functional forms of $\bm X^{\rm S}(\theta)$ and $\bm Z_0(\theta)$,
Eq.~\eqref{phase_model} provides a closed equation for the phases $\phi_i$
($i=1,\ldots, N$).

It is convenient to express $\Gamma_{ij}$ in terms of the Fourier
coefficients defined by
$U^{\rm S}(\theta) = \sum_{k=-\infty}^\infty u_k
\cos k \theta, Z_0^{(U)} (\theta) = \sum_{k=-\infty}^\infty -z_k \sin k \theta$, and $S_{ij}(\theta) =
\sum_{k=-\infty}^\infty s_k^{(ij)} \cos k \theta ~(u_k,z_k,s_k\in \mathbb R)$, where we assumed that 
$S_{ij}(\theta)$, $U^{\rm S}(\theta)$, and $Z_0^{(U)} (\theta)$ are
even, even, and odd functions, respectively. 
Substitute these expansions into Eq.~\eqref{gamma} with $\bm H_{ij}$
given by Eq.~\eqref{Hij}, we obtain a general expression: 
{
\setlength{\mathindent}{-2pt}
\begin{align}
 \Gamma_{ij} = & 2\pi \sum_{k,l} z_k u_l
  \left[ (-1)^l s_{l-k}^{(ij)}  \sin\left\{ (k+l) \eta_{ij} -k\phi_i
				       - l \phi_j \right\} \right .\nonumber \\
  &\left. - s_{-k-l}^{(ij)} \sin\left\{ (k+l) (\eta_{ij} -\phi_i)\right\}
				   				  \right].
\label{gamma_general}
\end{align}
}
For a regular hexagonal cell shape, we have
$s_k^{(ij)}=\frac{1}{k\pi}\sin \frac{k d_{ij}}{2}~(k\neq 0),
s_0^{(ij)}=\frac{d_{ij}}{2\pi}$. The coefficients $u_k$ and $z_k$ are
obtained for a given model.

\begin{figure}[t]
 \begin{center}
\includegraphics[scale=1.00]{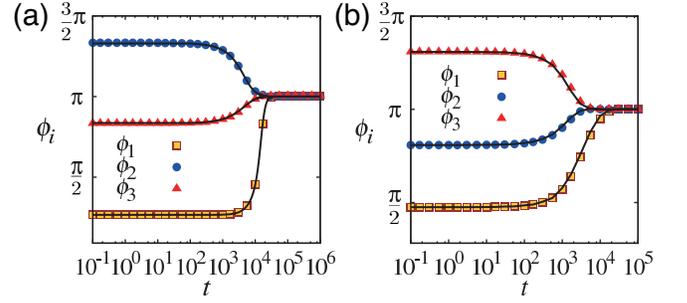}
 \end{center}
 \caption{(Color online). Comparison between the time series obtained from
 the reaction-diffusion
 models (symbols) and the corresponding phase models (lines).
 (a) GLE. (b) Ativator-inhibitor model.
 In this case, three hexagonal cells are aligned in a row, i.e.,
 $\eta_{12}=\eta_{23}=0, d_{12}=d_{23}=\frac{\pi}{3}$.
 }
   \label{fig:3cells}
\end{figure}

For the GLE, the phase reduction is analytically performed.
By solving $\bm F(\bm X^{\rm S}) + \hat D \frac{\partial^2}{\partial
\theta^2}\bm{X}^{\rm S}=0$, we obtain $\bm X^{\rm S} = (U^{\rm
S},V^{\rm S})= \sqrt{1-D_0}(\cos
\theta, \sin \theta)$.
By solving $\mathcal L^\dagger \bm Z_0 = 0$ with the
normalization $\langle \bm Z_0, \bm Y_0 \rangle = 1$,
where $\mathcal L^\dagger = \mathcal L$ in the
present model,
we obtain 
$\bm Z_0 = (Z_0^{(U)},Z_0^{(V)})=\frac{1}{2\pi\sqrt{1-D_0}} (\sin \theta, -\cos \theta)$ [\FIG{\ref{fig:X_Z}(a)}].
Therefore, Eq.~\eqref{gamma_general} reduces to
\begin{align}
 \Gamma_{ij}(\phi_i, \phi_j) =
 a_{ij}
 \sin(\phi_j-\phi_i) + \nonumber \\
 a_{ij} \sin2(\eta_{ij}-\phi_i)+ 
 b_{ij} \sin(2\eta_{ij}-\phi_i-\phi_j)
\label{gamma_cgle}
\end{align}
with $a_{ij} = \frac{\sin d_{ij}}{4\pi}$ and $b_{ij} = \frac{d_{ij}}{4\pi}$. 
For general models, phase reduction is performed numerically
by solving
Eq.~\eqref{noudo} for $\epsilon=0$
and its adjoint equation $\dot {\bm Z_0} = \mathcal L^\dagger \bm
Z_0$ with $\langle \bm Z_0, \bm Y_0 \rangle = 1$ \cite{kawamura15}.
For the activator-inhibitor model, 
$U^{\rm S}$ and $Z_0^{(U)}$ are obtained, as
shown in \FIG{\ref{fig:X_Z}(b)}. 
Their Fourier coefficients are approximately given as $u_0=0.925,
u_1=0.397, u_2=0.065, z_1=-0.180$, and $z_2=-0.062$. Other
coefficients are negligible in this case. For both the GLE
and the activator-inhibitor models,
the accuracy of our reduction theory if confirmed
by comparing the time series of the original model given by Eq.~\eqref{noudo}
and that of the phase model given by Eqs.~\eqref{phase_model} and \eqref{gamma_cgle}.
with the corresponding $\Gamma_{ij}$, as shown in \FIG{\ref{fig:3cells}}.

It should be noted that the phase sensitivity function $\bm Z_0(\theta)$ is
very useful for understanding the response of the polarity orientation to
perturbation. See Fig. \ref{fig:X_Z}(a) as
an example. If the $U$ variable is perturbed upward at $\theta=\pi/2$,
$\phi$ will increases because $\bm Z_0(\pi/2)>0$,
i.e., the pattern will eventually shift right.

{\em Analysis--.} 
We focus on the phase model with Eq.~\eqref{gamma_cgle} below because of
the following reason.
If $U^{\rm S}(\theta)$ and $Z_0^{(U)} (\theta)$ are nearly
harmonic, i.e., $u_k$ and $z_k$ ($k\ge 2$) are small, we approximately obtain
$ \Gamma_{ij} = - 4 \pi z_1 u_1
  [s_{2}^{(ij)} \sin(\phi_j -\phi_i)
   + s_{2}^{(ij)} \sin 2 (\eta_{ij} -\phi_i)
   + s_{0}^{(ij)} \sin (2 \eta_{ij} -\phi_i -  \phi_j )]$, which is 
Eq.~\eqref{gamma_cgle} with
generally different coefficients.
Therefore, the coupling function given by Eq.~\eqref{gamma_cgle}
is of crucial importance.

We first consider two coupled cells with $\eta_{12}=0, \eta_{21} = \pi,
a_{12}=a_{21}, b_{12}=b_{21}$,
and investigate the existence and stability of the in-phase state.
Substituting the in-phase state $(\phi_1, \phi_2) = (\phi^*,\phi^*)$
into
Eq.~\eqref{phase_model} with Eq.~\eqref{gamma_cgle},
we obtain $\sin 2\phi^* = 0$, thus $\phi^*=0, \pi$ or $\phi^*=\pm \frac{\pi}{2}$.
Putting $\phi_i = \phi^* + \psi_i$ ($i=1,2$), $\xi = \psi_1
+ \psi_2$ and $\zeta=\psi_1 - \psi_2$, and 
linearizing Eq.~\eqref{phase_model} for small
$\psi_i$, we obtain
\begin{align}
 \dot \xi &= -2 \epsilon (a_{ij}  + b_{ij}) (\cos 2\phi^*) \xi \\
 \dot \zeta &= -2 \epsilon a_{ij} ( 1 + \cos 2 \phi^*) \zeta.
\end{align}
The solutions $(0, 0)$ and $(\pi,\pi)$ are thus linearly stable when $\epsilon
a_{ij} > 0$ and $\epsilon (a_{ij}  + b_{ij}) > 0$.
The GLE with $\epsilon>0$ satisfies this condition.
In contrast, the solution $\phi=\pm\frac{\pi}{2}$ may not be
asymptotically stable
because $\dot \zeta$ always vanishes.
The same condition is obtained for the 1D straight chain of any number $N$ of
cells with open and periodic boundaries,
which can be shown by applying the Gershgorin circle theorem
to the corresponding stability matrix.

It should be emphasized that in Eq.~\eqref{gamma_cgle},
the second and third terms contain geometric information in
$a_{ij}, b_{ij}$ and $\eta_{ij}$ and they
facilitate the phase $\phi_i$ and
the mean phase $\frac{\phi_i +
\phi_j}{2}$ to be oriented to the cell-to-cell direction $\eta_{ij}$, respectively.
If only the first term is present in Eq.~\eqref{gamma_cgle},
which is the case in the XY model,
there is a family of stable solutions $(\phi_1, \phi_2) = (\phi^*,\phi^*)$
with arbitrary $\phi^*$ values, and
the realized polarity pattern is determined by
the initial conditions.
\begin{figure}[t]
 \includegraphics[scale=1.0]{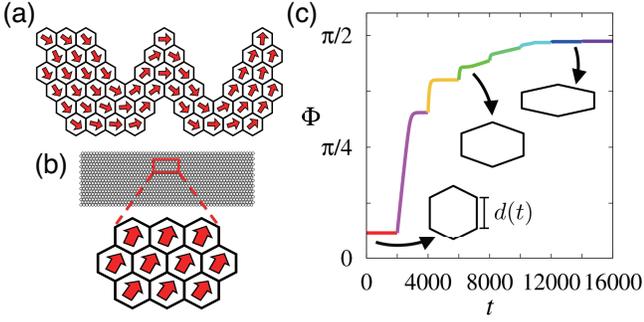}
 \caption{Polarity pattern for (a) winding cell alignment with a regular
 hexagonal shape and for (b,c) planar alignment of $60\times20$ cells with
 elongated shapes obtained numerically
 with the phase model [Eqs.~\eqref{phase_model} and
 \eqref{gamma_cgle}].
 In (a), the final pattern is displayed with each arrow
 indicating the phase of each cell.  In (b), phases at $t=3000$ are
 displayed.  In (c), the time series of the mean phase $\Phi(t)$ defined as $Q(t)
 e^{{\rm i} \Phi(t)} = \frac{1}{N}\sum_j e^{{\rm i} \phi_j(t)}$ with
 $Q \ge 0$ and $\Phi \in \mathbb R$ is displayed. In (b) and (c), cell shape
 is varied such that
 the regular hexagon is considered at $t=0$ and $d(t)$ [indicated
 in (c)] is decreased by $\frac{\pi}{30}$ at $t=2000 n$
 ($n=1,2,\ldots,7$), while
 keeping the perimeter $2\pi$.
 Initial conditions were chosen such that no topological defects appeared.
 }
 \label{fig:polarity}
\end{figure}

To obtain useful insight into dynamical behavior for a complicated
alignment of cells, we further simplify the phase model using the
assumption that the neighboring cells are nearly in phase.
Under the approximation that $\phi_i = \phi_j$ for any neighboring cells,
Eq.~\eqref{phase_model} with Eq.~\eqref{gamma_cgle} reduces to
\begin{equation}
 \dot \phi_i = \epsilon R_i \sin2(\overline {\eta_i} - \phi_i),
  \label{dot_phi_order}
\end{equation}
where $R_i>0$ and
$\overline{\eta_i} \in \mathbb R$
are determined by
$R_i e^{{\rm i} 2\overline{\eta_i}} = \sum_{j \in A(i)} (a_{ij} +
b_{ij}) e^{{\rm i} 2\eta_{ij}}$,
which can be interpreted as the effective strength and the preferred
direction of the net interaction of cell $i$, respectively.
We first consider square and hexagonal lattices, where
the cell has a square and regular hexagonal shape, respectively.
In these cases, $R_i$ vanishes for cell $i$ not
facing boundaries of the lattice because
$a_{ij}$ and $b_{ij}$ are not $i,j$-dependent and
$\eta_{ij}$ takes the values $0, 2\pi/n, 4\pi/n, \cdots, 2(n-1)\pi/n$ with $n=4$ and $6$
for the square and hexagonal lattices, respectively.
On the other hand, for cells at the boundary, $R_i$ is non-vanishing and
$\overline {\eta_i}$ is approximately parallel to the boundary line.
Therefore, cell polarity at the boundary is oriented parallel to the
boundary line
and the bulk is smoothly aligned to that of neighboring cells.
As shown in \FIG{\ref{fig:polarity}}(a),
this prediction is confirmed using the system with winding cell alignment.
In contrast, when the cell shape is elongated, $R_i$ is non-vanishing even
in the bulk. In this case, $\overline{\eta_i}$ tends to orient to
the direction of a contact surface with a larger width.
When the number of bulk units is much more than
that of boundary units, polarity orientation is dominantly dependent on the cell
shape. For example, as shown in \FIG{\ref{fig:polarity}(b,c)},
the polarity tends to be oriented to the direction of
the short axis as hexagonal cells are further elongated.

\begin{figure}[t]
\includegraphics[scale=1.0]{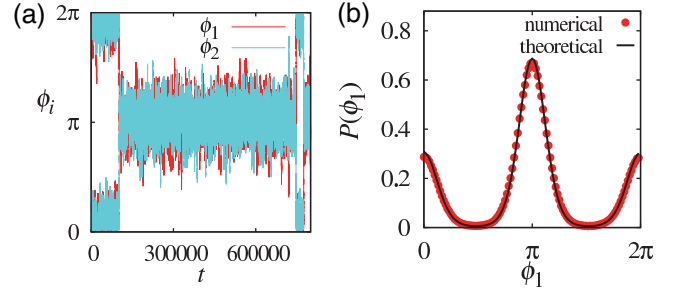}
 \caption{Polarity orientation in the presence of an external
 signal and noise in the GLE $(N=2)$. 
 (a) Time series obtained numerically from the reaction-diffusion model
 [Eq.~\eqref{noudo}]. (b) The probability density,
 where $P(\phi_1) = \int_0^{2\pi} P(\phi_1, \phi_2) d \phi_2$. 
 Numerical results are obtained from direct simulation of Eq.~\eqref{noudo} with
 an inclusion of additive noise $\bm p_i(\theta_i, t) = (p_i^{(1)}, 0)$
 and external signal $\bm G_i = (\cos(\psi-\theta_i),0)$ with $\psi=\pi$. 
 (a) Time series. (b) The probability density function obtained numerically
 and the theoretical one $P(\phi_1) = \int_0^{2\pi}
 P(\phi_1, \phi_2) d \phi_2$. The parameter values were $\eta_{12}=0,
 \nu_1=0.005$, and $D_0=0.2$ (only for this figure).
 }
 \label{fig:noise}
\end{figure}

The phase reduction is also possible 
when our reaction-diffusion model includes external signal and noise
(see Supplemental Material B for details).
Specifically, we add to Eq.~\eqref{noudo}
external signal $\eps_{\rm e} \bm G_i(\theta_i,t)$ and
white Gaussian noise $\bm p_i(\theta_i, t) = (p_i^{(1)}, p_i^{(2)},
\ldots)$ that satisfies 
${\rm E}[p_i^{(m)}] =0$ and ${\rm E}[p_i^{(m)} (\theta_i, t) p_j^{(n)} (\theta_j, t')] =
\nu_m \delta_{ij} \delta_{mn} \delta(\theta_i-\theta_j) \delta(t-t')$ where
${\rm E}[\cdot]$ denotes the ensemble average and $\nu_m$ is the noise
intensity. For sufficiently small $\eps_{\rm e}$ and $\nu_m$,
we obtain 
\begin{align}
 \dot \phi_i = \eps \sum_{j\in A(i)} \Gamma_{ij}(\phi_i, \phi_j) +
 \eps_{\rm e} \Pi_i(\phi_i)
 + q_i,
 \label{phase_model_noisy}
\end{align}
where $\Pi_i(\phi_i) =  \langle \bm Z_0 (\theta_i-\phi_i), \bm G_{i}(\theta_i) \rangle$
and $q_i(t)$ is a white Gaussian noise with
zero mean and variance 
$\nu = \sum_m \nu_m \int_0^{2\pi}  {Z^{(m)} (\theta)}^2 d \theta$ with 
$Z^{(m)}$ being the $m$th component of $\bm Z$.
As a simple example, we consider the GLE with
$\bm G_i = (\cos(\psi-\theta_i),0)$, where $\psi$ is a parameter,
resulting in $\Pi_i = \frac{1}{2\sqrt{1-D_0}}\sin(\psi-\theta_i)$.
The phase model under consideration is
actually a gradient system, i.e.,
$\dot \phi_i = -\frac{\partial}{\partial \phi_i} \mathcal H  + q_i$
with the potential function $\mathcal H = \mathcal H(\{\phi_i\})$ given
in Supplemental Material B for details.
We thus obtain probability distribution
$P(\{\phi_i\}) = C \exp\left[ -\frac{2\epsilon\mathcal
H(\{\phi_i\})}{\nu} \right]$,
where $C$ is the normalization constant.
As shown in \FIG{~\ref{fig:noise}}, the probability distribution obtained
numerically from the reaction-diffusion model, Eq.~\eqref{noudo}, is in
excellent agreement with $P(\{\phi_i\})$.

{\em Discussion and Conclusion--}.
A theoretical framework
for understanding dynamical properties of alignment process of
cellular polarity was proposed.
Although the phenomena of our concern are highly nonlinear,
our framework enables their analytical treatment even in the presence of
noise.
Our described framework
is readily extendable
to treat more concrete problems.
For example, the effects of cell heterogeneity 
and cell shape dependence on local cellular dynamics were examined
in previous studies on PCP \cite{amonlirdviman05,aigouy10}. These
factors can be incorporated into our reaction-diffusion model and the resulting phase
model and its dynamical behavior would be of great interest.
Overall, we expect that our framework would have many potential applications
in nonequilibrium systems including chemical and biological systems.

\begin{acknowledgements}
We are grateful to Dr. Masakazu Akiyama,
Dr. Hugues Chate, Dr. Yasuaki Kobayashi,
Dr. Yoji Kawamura, Dr. Hiroya Nakao, and Dr. Alexander Mikhailov
for helpful discussion and comments.
We acknowledge the financial support from CREST, JST and JSPS KAKENHI
Grant No. 15K16062.
\end{acknowledgements}

\clearpage

\onecolumngrid
\appendix
\setcounter{page}{1}
\renewcommand\appendixname{Supplemental Material}%

\begin{center}
 {\Large Supplemental Material for\\
 ``A Simple Generic Model of Cellular Polarity Alignment:\\
 Derivation and Analysis''}
 \vspace{1cm}

 K. Sugimura$^1$ and H. Kori$^1$

{\it $^1$Department of Information Sciences, Ochanomizu University, Tokyo
 112-8610, Japan.}

\end{center}

\makeatletter
    \renewcommand{\thefigure}{%
    S\arabic{figure}}
\makeatother

\makeatletter
    \renewcommand{\theequation}{%
    S\arabic{equation}}
 \makeatother

\setcounter{figure}{0}
\setcounter{equation}{0}

\section{Model equations} \label{appendix:model}
Our reaction-diffusion model in the absence of perturbation is given as
\begin{align}
 \frac{\partial}{\partial t} \bm{X}_{i}
  &= \bm{F}(\bm{X}_{i})
  + \hat D \frac{\partial^2 \bm{X}_{i}}{\partial \theta_i^2}.
\end{align}
We consider two exmple models: (a)the real Ginzburg-Landau equation
(GLE) and (b)the activator-inhibitor model. 
With $\bm X_i=(U_i,V_i)$, 
the former reads
\begin{align}
 \bm F =\left(
\begin{array}{c}
 U_i-({U_i}^2+{V_i}^2) U_i\\
 V_i-({U_i}^2+{V_i}^2)V_i)
\end{array}
 \right),
\end{align}
where $\hat D={\rm diag}(D_0, D_0)$ and $D_0=0.3$.
The latter reads
\begin{align}
  \bm F = \left(
 \begin{array}{c}
  \frac{\rho_U U_i^2}{(1+\kappa U_i^2)V_i} - \mu_U U_i + \sigma_U \\
  \rho_V U_i^2 - \mu_V V_i,
 \end{array} 
 \right),
\end{align}
where $\rho_U = 0.01, \rho_V = 0.02, \mu_U = 0.01, \mu_V = 0.02, \sigma_U
= 0.0, \kappa = 0.0, \hat D={\rm diag}(D_U, D_V), D_U=0.005, D_V = 0.2$, respectively.

\section{Phase reduction in the presence of external signal and noise} \label{appendix:noise}
Our reaction-diffusion model
in the presense of intercellular interaction, external signal, and noise is given as
\begin{eqnarray}
 \frac{\partial}{\partial t} \bm{X}_{i}
  = \bm{F}(\bm{X}_{i})
  + \hat D \frac{\partial^2}{\partial \theta_i^2}\bm{X}_{i}
  + \eps \sum_{j\in A(i)} \bm H_{ij} + \eps_{\rm e} \bm G_i +
  \bm p_i,
\label{noudo_stochastic}
\end{eqnarray}
where $\bm G_i(\theta_i,t)$ is the external signal, $\eps_{\rm e}$ is its
strength, and $\bm p_i = (p_i^{(1)}, p_i^{(2)},\ldots)$ is white Gaussian noise
that satisfies 
${\rm E}[p_i^{(m)}(\theta, t)] =0$
and ${\rm E}[p_i^{(m)} (\theta, t) p_j^{(n)} (\theta', t')] =
\nu_m \delta_{ij} \delta_{mn} \delta(\theta-\theta') \delta(t-t')$,
and $\nu_m$ is the noise intensity.
For sufficiently small $\eps_{\rm e}$ and $\nu_m$,
we carry on the same procedure as that for Eq.~\eqref{noudo} to 
obtain 
\begin{align}
 \dot \phi_i = \epsilon \sum_{j\in A(i)} \Gamma_{ij}(\phi_i, \phi_j)
 + \eps_{\rm e} \Pi_i(\phi_i,t) + q_i(t)
 \label{phase_model_noisy_appendix}
\end{align}
where
\begin{align}
 \Pi_i(\phi_i) &=  \langle \bm Z_0 (\theta_i-\phi_i), \bm G_{i}(\theta_i,t)
 \rangle, \\
 q_i(t) &=  \langle \bm Z_0(\theta_i-\phi_i), \bm p_i(\theta_i, t)\rangle.
\end{align}
Note that $q_i(t)$ is Gaussian white noise that
satisfies ${\rm E}[q_i(t)]=0$ and ${\rm E}[q_i(t) q_j(t')]=
\nu \delta_{ij} \delta(t-t')$ with $\nu = \sum_m \nu_m \int_0^{2\pi}
d\theta \left\{Z^{(m)}(\theta)\right\}^2$ because
\begin{align}
 {\rm E}[q_i(t)] &= {\rm E}\left[\int_0^{2\pi} d\theta
 Z(\theta-\phi_i) \cdot \bm p_i(t) d\theta \right] \\
 &={\rm E}\left[\int_0^{2\pi} d\theta
 \sum_m Z^{(m)}(\theta-\phi_i)p_i^{(m)} d\theta \right] \\
 &=\int_0^{2\pi} d\theta \sum_m Z^{(m)}(\theta-\phi_i) {\rm
 E}[p_i^{(m)}] d\theta\\
 &=0,
\end{align}
and
\begin{align}
 {\rm E}[q_i(t) q_j(t')] &= {\rm E}\left[\iint_0^{2\pi} d\theta d\theta' \{\bm
 Z(\theta-\phi_i(t)) \cdot \bm p_i(t)\}\{\bm Z(\theta'-\phi_j(t')) \cdot \bm p_j(t')\}\right]\\
 &= {\rm E}\left[ \iint_0^{2\pi}\int_0^{2\pi} d\theta d\theta'
 \left\{\sum_m Z^{(m)}(\theta-\phi_i(t))p_i^{(m)}(t)\right\}
 \left\{\sum_{m'} Z^{(m')}(\theta'-\phi_j(t'))p_j^{(m')}(t')\right\}\right]\\
 &= {\rm E}\left[ \iint_0^{2\pi} d\theta d\theta'
 \sum_{m,m'} Z^{(m)}(\theta-\phi_i(t))p_i^{(m)}
 Z^{(m')}(\theta'-\phi_j(t'))p_j^{(m')}(t') \right]\\
 &= \iint_0^{2\pi} d\theta d\theta'
 \sum_{m, m'} Z^{(m)}(\theta-\phi_i(t)) Z^{(m')}(\theta'-\phi_j(t'))
 {\rm E}\left[ p_i^{(m)}(t) p_j^{(m')}(t') \right]\\
 &= \iint_0^{2\pi} d\theta d\theta'
 \sum_{m,m'} Z^{(m)}(\theta-\phi_i(t)) Z^{(m')}(\theta'-\phi_j(t'))
 \nu_m \delta_{ij} \delta_{mm'} \delta(\theta-\theta') \delta(t-t') \\
 &= \int_0^{2\pi} d\theta
 \sum_m \nu_m \left\{Z^{(m)}(\theta-\phi_i(t))\right\}^2\\
 &=  \sum_m \nu_m \int_0^{2\pi} d\theta \left\{Z^{(m)}(\theta)\right\}^2.
\end{align}

In the case of GLE, any generic choice of external signal $\bm G_i(\theta_i,t)$
yields 
\begin{equation}
 \Pi_i = c_i(t) \sin(\psi_i(t) - \theta_i)
  \label{Pi}
\end{equation}
because $\bm Z_0(\theta)$ contains only the first harmonics.
As a simple example, we consider
\begin{equation}
 \bm G_i(\theta_i) = (\cos(\psi (t) -\theta_i),0),
\end{equation}
where $\psi_i(t)$ is a parameter, we obtain
\begin{equation}
 \Pi_i = \frac{1}{2\sqrt{1-D_0}}\sin(\psi-\theta_i).
\end{equation}
\end{document}